\documentclass[10pt]{article}
\usepackage{amsmath,amsfonts,amssymb,amsthm,latexsym,graphicx,textcomp}

\allowdisplaybreaks[4]

\newcommand{\st}{\mathrel{|}}

\DeclareMathOperator{\Sh}{\textrm{Sh}}	
\DeclareMathOperator{\So}{\textrm{So}}	

\theoremstyle{plain}

\newtheorem*{problem*}{Problem}

\theoremstyle{definition}

\title{Losing money with a high Sharpe ratio}
\author{Vladimir Vovk}

\begin{document}
\maketitle

\begin{abstract}
  A simple example shows that losing all money
  is compatible with a very high Sharpe ratio
  (as computed after losing all money).
  However, the only way that the Sharpe ratio can be high
  while losing money is that there is a period
  in which all or almost all money is lost.
  This note explores the best achievable Sharpe and Sortino ratios
  for investors who lose money
  but whose one-period returns are bounded below
  (or both below and above) by a known constant.
\end{abstract}

\section{Introduction}

Sharpe ratio \cite{sharpe:1966,sharpe:1994} has become a ``gold standard''
for measuring performance of hedge funds and other institutional investors
(this note uses the generic term ``portfolio'').
It is sometimes argued that it is applicable only to i.i.d.\ Gaussian returns,
but we will follow a common practice of ignoring such assumptions.
For simplicity we assume that the benchmark return
(such as the risk-free rate) is zero.

The (ex post) \emph{Sharpe ratio} of a sequence of returns $x_1,\ldots,x_N\in[-1,\infty)$
is defined as $\Sh_N=\Sh(x_1,\ldots,x_N):=\mu_N/\sigma_N$,
where
$$
  \mu_N:=\frac1N\sum_{n=1}^Nx_n,
  \quad
  \sigma_N:=\sqrt{\frac{1}{N}\sum_{n=1}^N(x_n-\mu)^2}.
$$
(None of our results will be affected if we replace,
assuming $N\ge2$,
$\frac1N$ by $\frac{1}{N-1}$,
as in \cite{sharpe:1994}, (6).)
Intuitively, the Sharpe ratio is the return per unit of risk.

Another way of measuring the performance of a portfolio
whose sequence of returns is $x_1,\ldots,x_N$
is to see how this sequence of returns would have affected an initial investment of 1
assuming no capital inflows and outflows after the initial investment.
The final capital resulting from this sequence of returns is
$\prod_{n=1}^N(1+x_n)$.
We are interested in conditions under which the following anomaly is possible:
the Sharpe ratio $\Sh(x_1,\ldots,x_N)$ is large while $\prod_{n=1}^N(1+x_n)<1$.
(More generally, if we did not assume zero benchmark returns,
we would replace $\prod_{n=1}^N(1+x_n)<1$
by the condition that in the absence of capital inflows and outflows
the returns $x_1,\ldots,x_N$ underperform the benchmark portfolio.)

Suppose the return is $5\%$ over $k-1$ periods,
and then it is $-100\%$ in the $k$th period.
As $k\to\infty$, $\mu_k\to0.05$ and $\sigma_k\to0$.
Therefore, making $k$ large enough,
we can make the Sharpe ratio $\Sh_k$ as large as we want,
despite losing all the money over the $k$ periods.

If we want the sequence of returns to be i.i.d.,
let the return in each period $n=1,2,\ldots$
be $5\%$ with probability $(k-1)/k$ and $-100\%$ with probability $1/k$,
for a large enough $k$.
With probability one the Sharpe ratio $\Sh_N$ will tend to a large number as $N\to\infty$,
despite all money being regularly lost.
Of course, in this example the returns are far from being Gaussian
(strictly speaking, returns cannot be Gaussian unless they are constant,
since they are bounded from below by $-1$).

It is easy to see that our examples lead to the same conclusions
when the Sharpe ration is replaced by the \emph{Sortino ratio}
\cite{sortino/meer:1991,sortino/price:1994}
$\So_N=\So(x_1,\ldots,x_N):=\mu_N/\sigma'_N$,
where
$$
  \sigma'_N:=\sqrt{\frac{1}{N}\sum_{n=1}^N((x_n-\mu)^-)^2}\le\sigma_N.
$$

\section{Upper bound on the Sharpe ratio for losers}

The examples of the previous section are somewhat unrealistic
in that there is a period in which the portfolio loses
almost all its money.
In this section we show that only in this way
a high Sharpe ratio can become compatible with losing money.

For each $B\in(0,1]$,
define
\begin{equation}\label{eq:F1}
  F_1(B)
  :=
  \sup
  \left\{
    \Sh(x_1,\ldots,x_N)
    \st
    \prod_{n=1}^N (1+x_n) < 1
  \right\},
\end{equation}
where $N$ ranges over the positive integers
and $(x_1,\ldots,x_N)$ over $[-B,\infty)^N$.
In other words, $F_1(B)$ is the best achievable Sharpe ratio
for sequences of returns that lose money,
assuming that none of the returns falls below $-B$.

It is not difficult to show that $F_1(0+)=0$,
and in the previous section we saw that $F_1(1)=\infty$.
In this section we are interested in the behaviour of $F_1(B)$
for the intermediate values of $B$, $B\in(0,1)$.

\begin{figure}
  \begin{center}
    \includegraphics[width=0.48\textwidth]{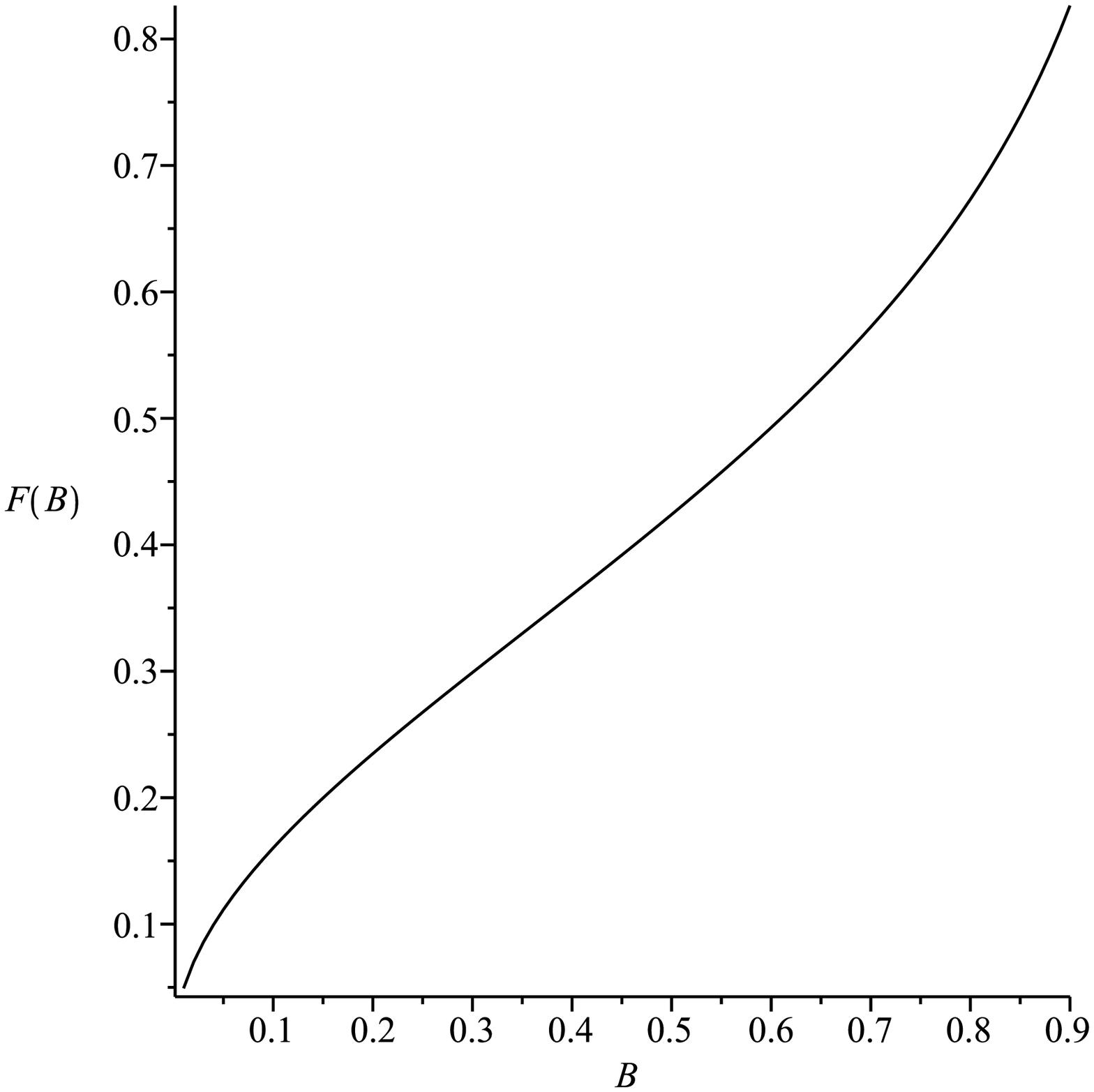}
    \includegraphics[width=0.48\textwidth]{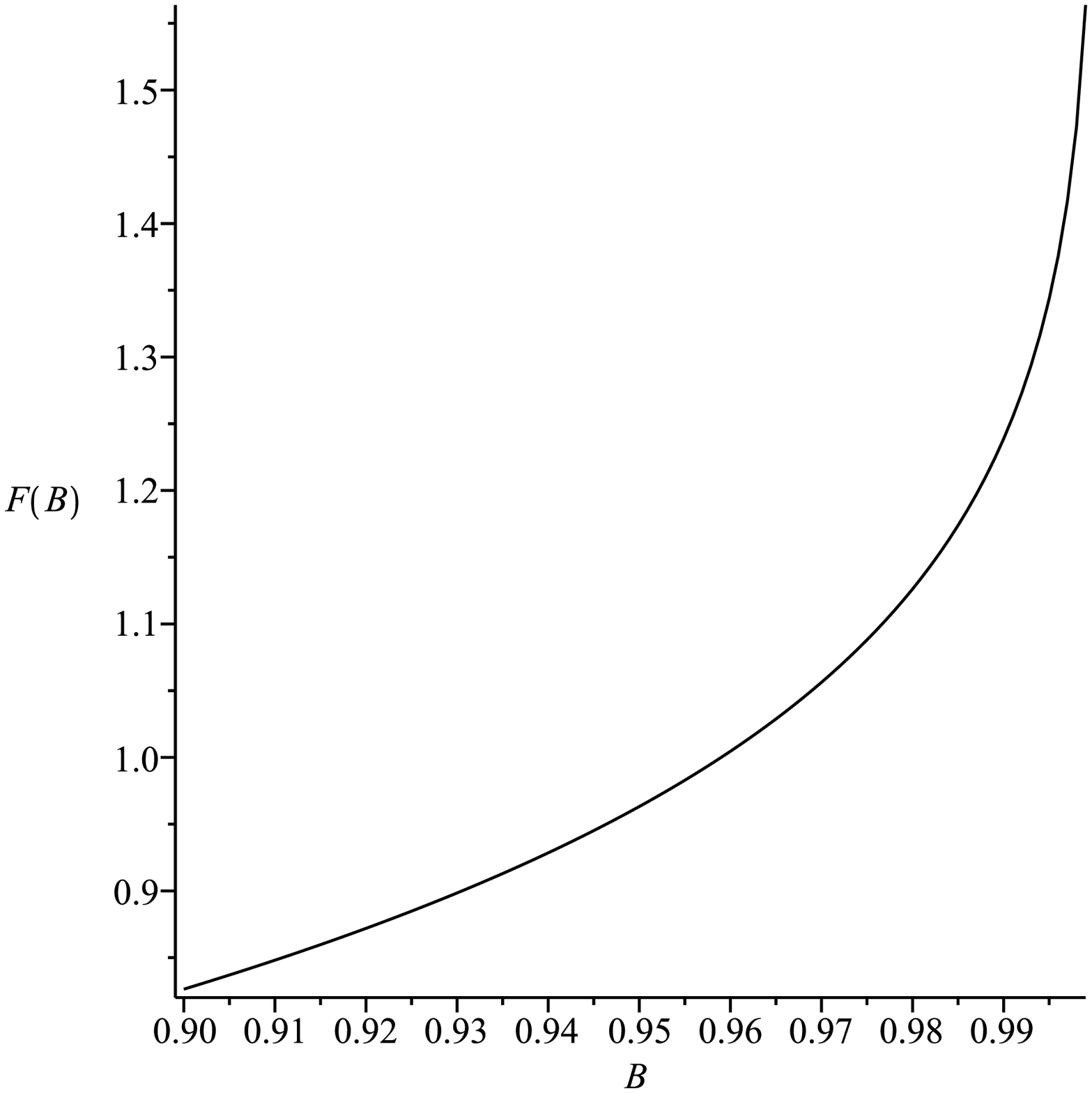}
  \end{center}
  \caption{\label{fig:F1}
    The function $F_1(B)$ in the ranges $B\in(0,0.9]$ (left)
    and $B\in[0.9,0.999]$ (right).}
\end{figure}

Figure~\ref{fig:F1} shows the graph of $F_1$ over $B\in(0,0.9]$
and over $B\in[0.9,0.999]$.
Over the interval $B\in(0,0.9]$ the slope of $F_1$ is roughly 1.
We can see that even for a relatively large value of $B=0.5$,
the Sharpe ratio of a losing portfolio never exceeds 0.5;
according to Table~\ref{tab:F1}, $F_1(0.5)=0.424$
(much less than the conventional threshold of 1
for a good Sharpe ratio \cite{investopedia:2010}).

\begin{table}
\begin{center}
\begin{tabular}{|c|c|c|c||c|c|c|c|}
\hline
$B$ & $F_1(B)$ & $c$ & $\alpha$ & $B$ & $F_1(B)$ & $c$ & $\alpha$\\
\hline
0.1 & 0.160 & 10.49 & 0.041  &  0.8 & 0.673 & 6.37 & 0.446\\
0.2 & 0.235 & 10.00 & 0.085  &  0.9 & 0.826 & 5.42 & 0.553\\
0.3 & 0.299 & 9.50 & 0.132   &  0.99 & 1.239 & 3.90 & 0.743\\
0.4 & 0.361 & 8.97 & 0.182   &  0.999 & 1.564 & 3.35 & 0.824\\
0.5 & 0.424 & 8.40 & 0.236   &  0.9999 & 1.836 & 3.10 & 0.867\\
0.6 & 0.493 & 7.80 & 0.296   &  0.99999 & 2.075 & 2.96 & 0.893\\
0.7 & 0.572 & 7.13 & 0.365   &  0.999999 & 2.289 & 2.88 & 0.911\\
\hline
\end{tabular}
\end{center}
\caption{\label{tab:F1}The approximate values of $F_1(B)$, $c$, and $\alpha$
  for selected $B$.}
\end{table}

Table~\ref{tab:F1} gives approximate numerical values of $F_1(B)$ for selected $B$.
These approximate values are attained for sequences of returns
involving only two levels of returns: $-B$ and another level $c$.
The table also lists the value of $c$
and the fraction $\alpha$ of returns equal to $c$
at which the given approximate value of $F_1(B)$ is attained.

\begin{figure}
  \begin{center}
    \includegraphics[width=0.48\textwidth]{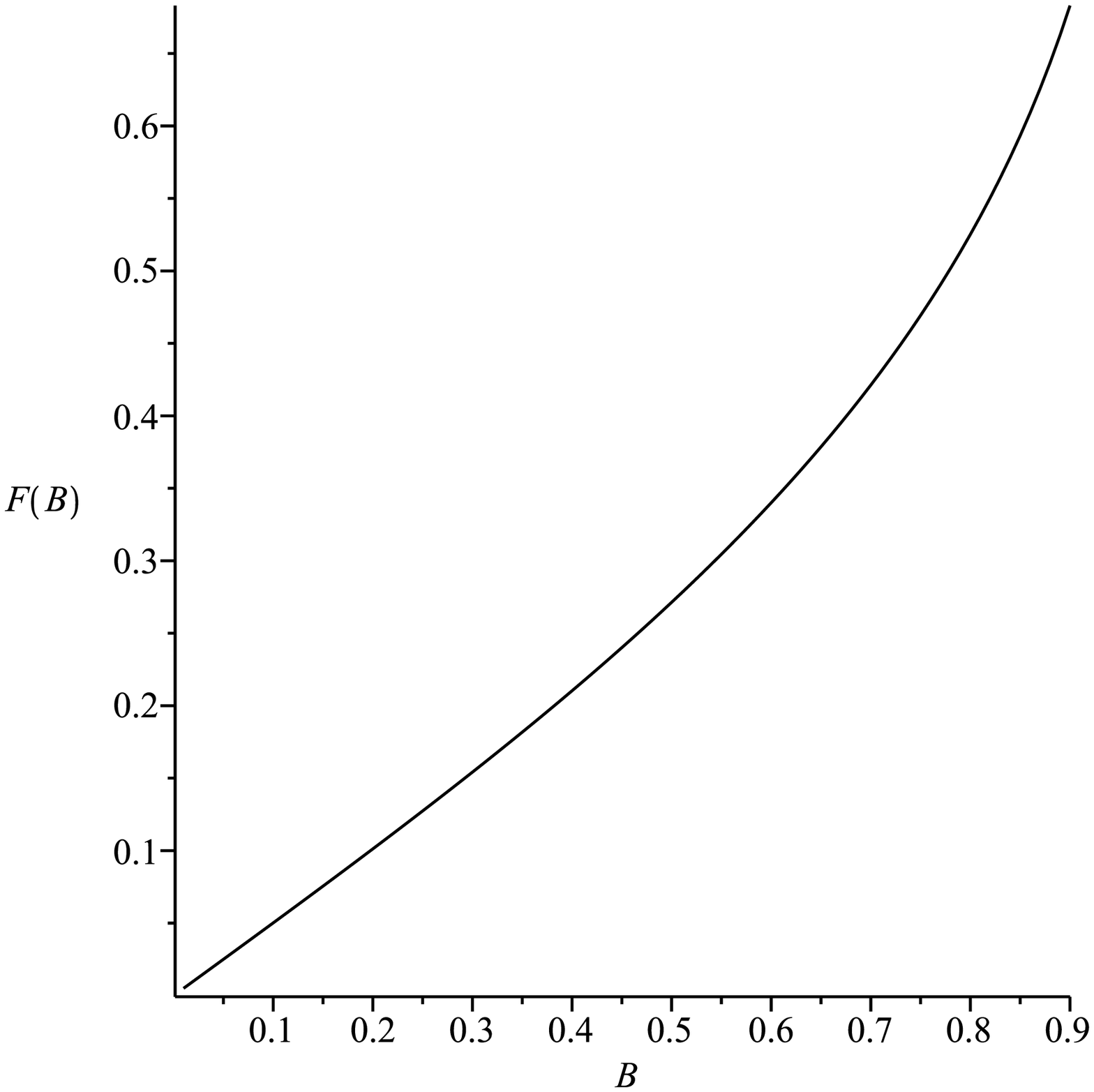}
    \includegraphics[width=0.48\textwidth]{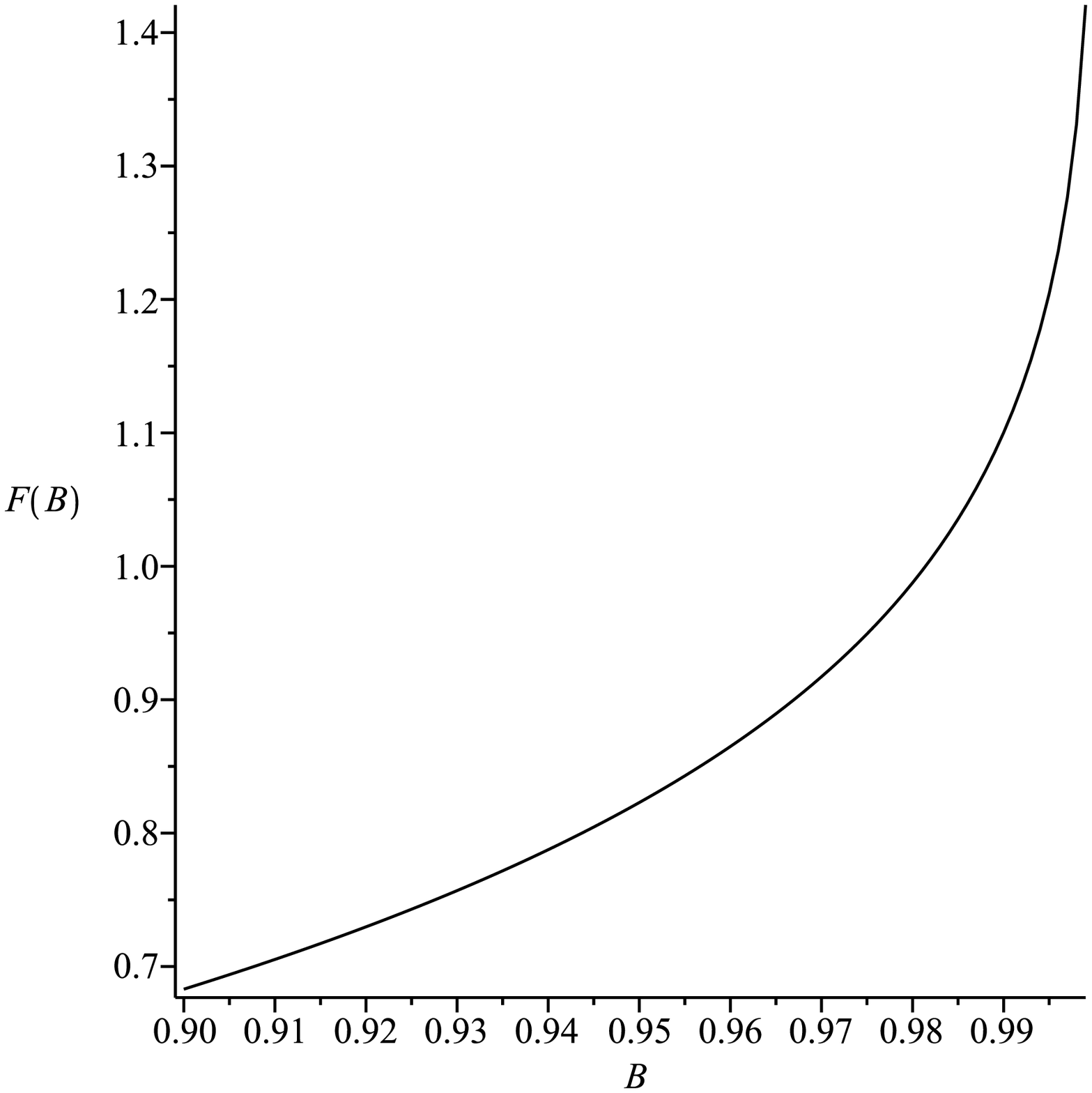}
  \end{center}
  \caption{\label{fig:F2}
    The function $F_2(B)$ in the ranges $B\in(0,0.9]$ (left)
    and $B\in[0.9,0.999]$ (right).}
\end{figure}

\begin{table}
\begin{center}
\begin{tabular}{|c|c|c||c|c|c|}
\hline
$B$ & $F_2(B)$ & $\alpha$ & $B$ & $F_2(B)$ & $\alpha$\\
\hline
0.1 & 0.050 & 0.525  &  0.4 & 0.210 & 0.603\\
0.2 & 0.101 & 0.550  &  0.5 & 0.271 & 0.631\\
0.3 & 0.154 & 0.576  &  0.6 & 0.340 & 0.661\\
\hline
\end{tabular}
\end{center}
\caption{\label{tab:F2}The approximate values of $F_2(B)$ and $\alpha$
  for selected $B$.}
\end{table}

A striking feature of Table~\ref{tab:F1} is the values of $c$:
they exceed $1$ even for $B=0.999999$.
The value of $c=10.00$ corresponding to $B=0.2$ in Table~\ref{tab:F1}
means that the portfolio increases its value 11-fold in one period.
Figure~\ref{fig:F2} is analogous to Figure~\ref{fig:F1}
but imposes the upper bound of $B$ on the absolute values of one-period returns.
Namely, it plots the graph of the function $F_2$
which is defined by the same formula (\ref{eq:F1}) as $F_1$
but with $(x_1,\ldots,x_N)$ now ranging over $[-B,B]^N$.
An abridged analogue of Table~\ref{tab:F1} for $F_2$
is given as Table~\ref{tab:F2};
the latter does not give the value of $c$ as it is equal to $B$ in all the entries.
Not surprisingly $\alpha$ are not so different from $1/2$ in Table~\ref{tab:F2},
especially for smaller $B$.

\begin{figure}
  \begin{center}
    \includegraphics[width=0.48\textwidth]{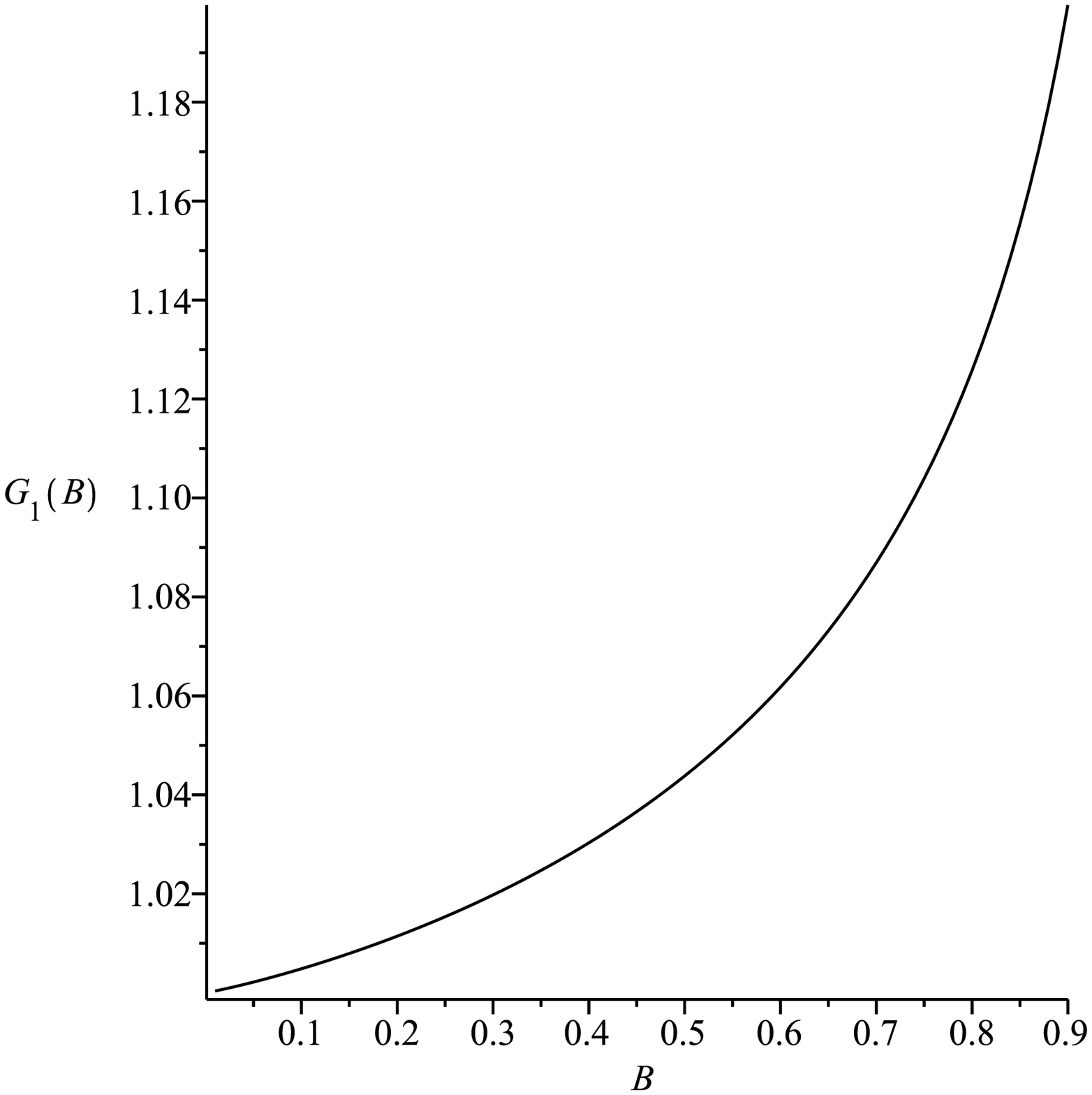}
    \includegraphics[width=0.48\textwidth]{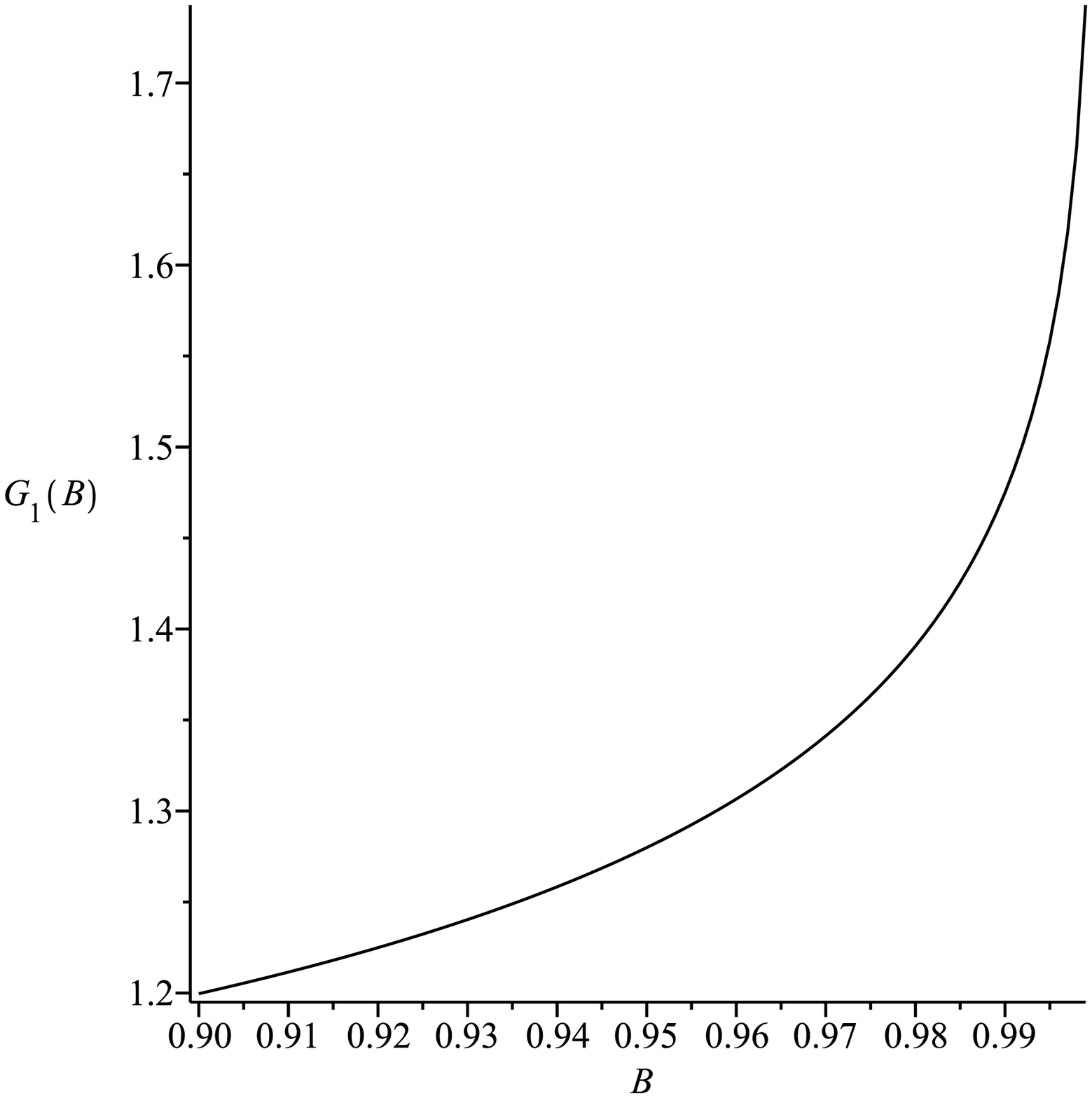}
  \end{center}
  \caption{\label{fig:G1}
    The function $G_1(B)$ in the ranges $B\in(0,0.9]$ (left)
    and $B\in[0.9,0.999]$ (right).}
\end{figure}

\begin{figure}
  \begin{center}
    \includegraphics[width=0.48\textwidth]{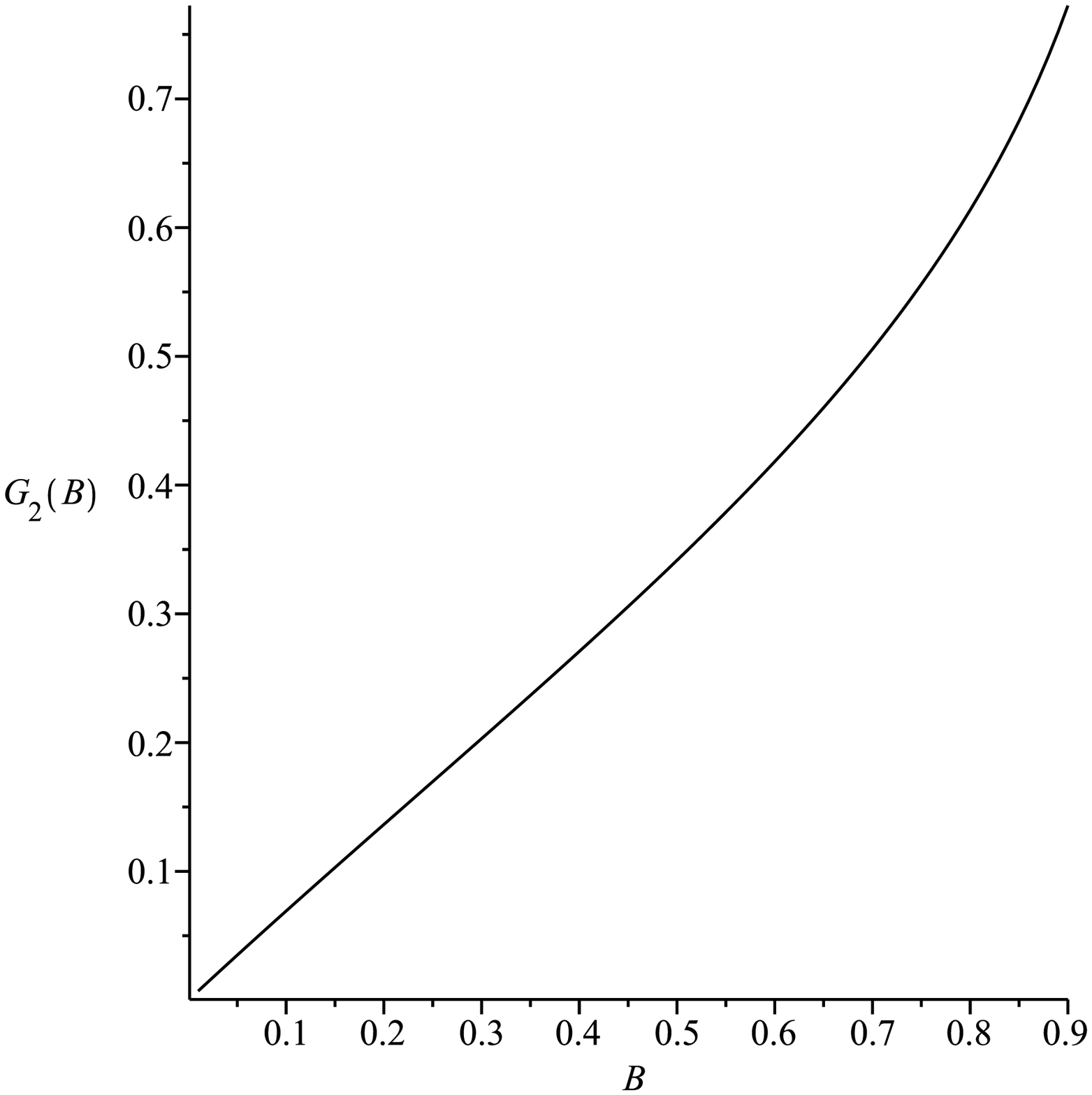}
    \includegraphics[width=0.48\textwidth]{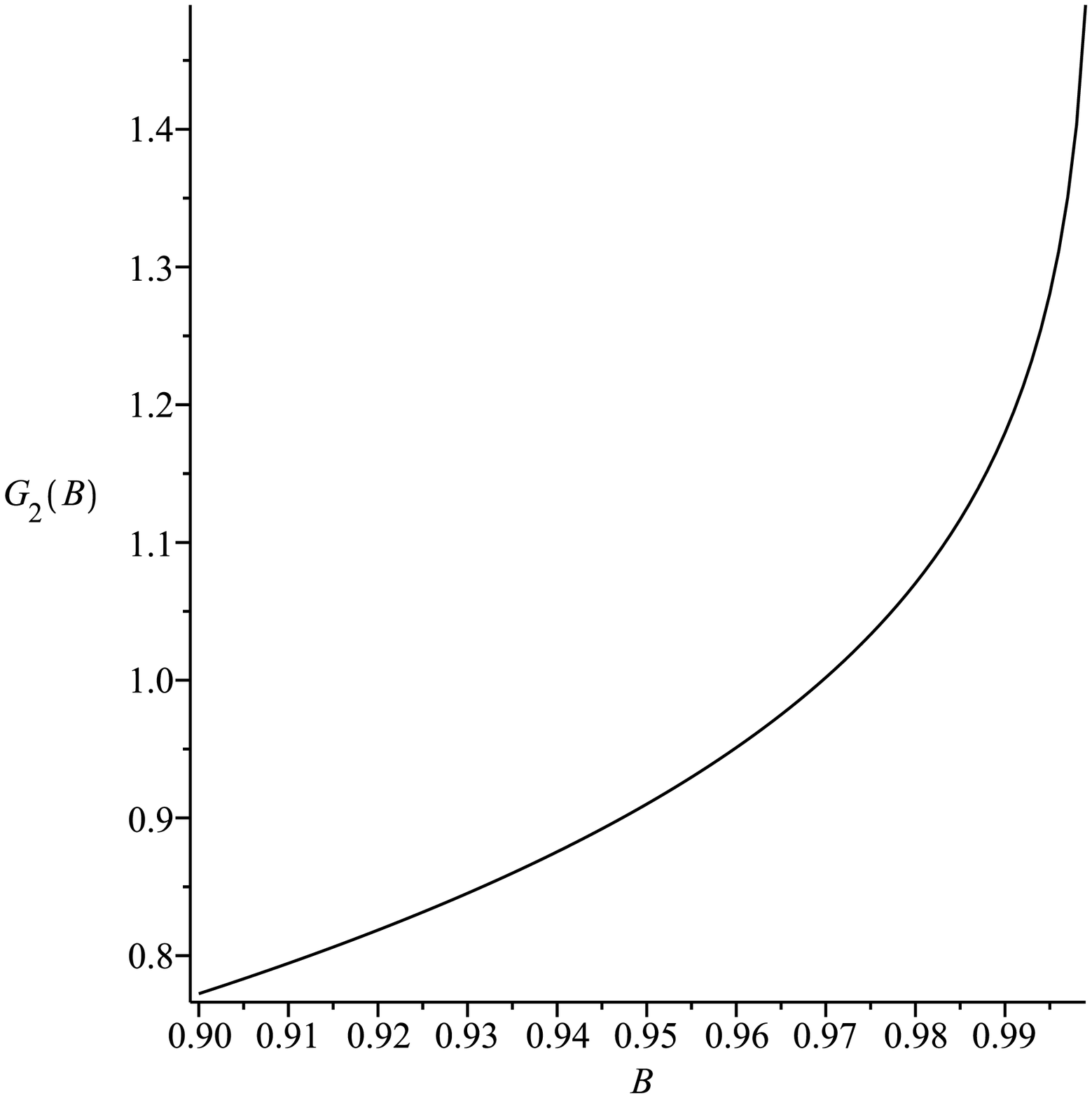}
  \end{center}
  \caption{\label{fig:G2}
    The function $G_2(B)$ in the ranges $B\in(0,0.9]$ (left)
    and $B\in[0.9,0.999]$ (right).}
\end{figure}

The analogues of Figures~\ref{fig:F1} and~\ref{fig:F2}
for the Sortino ratio are Figures~\ref{fig:G1} (one-sided)
and~\ref{fig:G2} (two-sided).
The function $G_1$ of Figure~\ref{fig:G1}
is defined, in analogy with (\ref{eq:F1}), by
\begin{equation}\label{eq:G1}
  G_1(B)
  :=
  \sup
  \left\{
    \So(x_1,\ldots,x_N)
    \st
    \prod_{n=1}^N (1+x_n) < 1
  \right\},
\end{equation}
where $N$ ranges over the positive integers
and $(x_1,\ldots,x_N)$ over $[-B,\infty)^N$.
The function $G_2$ of Figure~\ref{fig:G2}
is defined as the right-hand side of (\ref{eq:G1})
but with $(x_1,\ldots,x_N)$ ranging over $[-B,B]^N$.

\begin{table}
\begin{center}
\begin{tabular}{|c|c|c||c|c|c|}
\hline
$B$ & $G_1(B)$ & $\alpha$ & $B$ & $G_2(B)$ & $\alpha$\\
\hline
0.1 & 1.005 & 0.011  &  0.1 & 0.069 & 0.525\\
0.2 & 1.011 & 0.026  &  0.2 & 0.136 & 0.550\\
0.3 & 1.020 & 0.045  &  0.3 & 0.203 & 0.576\\
0.4 & 1.030 & 0.069  &  0.4 & 0.271 & 0.603\\
0.5 & 1.044 & 0.099  &  0.5 & 0.341 & 0.631\\
0.6 & 1.062 & 0.137  &  0.6 & 0.418 & 0.661\\
\hline
\end{tabular}
\end{center}
\caption{\label{tab:G}The approximate values of $G_1(B)$ and $\alpha$ (left)
  and $G_2(B)$ and $\alpha$ (right) for selected $B$.}
\end{table}

The values of $G_1(B)$ and $G_2(B)$ for selected $B$
are shown in Table~\ref{tab:G},
$G_1$ on the left and $G_2$ on the right.
The meaning of $\alpha$ is the same as in Tables~\ref{tab:F1} and~\ref{tab:F2}.
We do not give the values of $c$;
they are huge on the left-hand side of the table
and equal to $B$ on the right-hand side.
The left-hand side suggests that $G_1(0+)=1$,
and this can be verified analytically.

\section{Discussion}

Figures \ref{fig:F1}--\ref{fig:G2} can be regarded as a sanity check
for the Sharpe and Sortino ratio.
Not surprisingly, they survive it,
despite the theoretical possibility of having a high Sharpe
and, \emph{a fortiori}, Sortino ratio while losing money.
In the case of the Sharpe ratio,
such an abnormal behaviour can happen only when some one-period returns
are very close to $-1$.
In the case of the Sortino ratio,
such an abnormal behaviour can happen only when some one-period returns
are very close to $-1$
or when some one-period returns are huge.

\subsection*{Acknowledgements}

I am grateful to Boris Afanasiev for his advice
and to Wouter Koolen for useful comments.
All computations for this note were done in Maple$^{\textrm{\texttrademark}}$.

\end{document}